\documentclass[%
 reprint,
 amsmath,amssymb,
 aps,
]{revtex4-2}

\usepackage{subeqnar}
\usepackage{graphicx}
\usepackage{dcolumn}
\usepackage{bm}

\begin{document}

\preprint{APS/123-QED}

\title{Spatio-Temporal Mode-locking in Quadratic Nonlinear Media}

\author{Mahmut Ba\u gc\i}
\email{bagcimahmut@gmail.com}
\affiliation{
  Department of Computer Technology, Istanbul Bilgi University,  Kozyatagi 34742, Istanbul, Turkey \\
}%


\author{J. Nathan Kutz}
\affiliation{Department of Applied Mathematics, University of Washington, Seattle, WA 98195-2420 USA\\
}%


\date{\today}
\begin{abstract}
 A new theoretical model is developed to characterize spatio-temporal mode-locking (ML) in quadratic nonlinear media. The model is based on the two-dimensional nonlinear Schr\"odinger equation with coupling to a mean term (NLSM) and constructed as an extension of the master mode-locking model. It is numerically demonstrated that there exists steady state soliton solutions of the ML-NLSM mode-locking model that are astigmatic in nature. A full stability analysis and bifurcation study is performed for the ML-NLSM model and it is manifest that spatio-temporal mode-locking of the astigmatic steady-state solutions is possible in quadratic nonlinear media.
\end{abstract}

\maketitle


\section{\label{sec:level1} Introduction}

Mode-locking (ML) is a commonly observed phenomenon in optical resonator cavities where nonlinear interactions in the cavity synchronize different cavity modes to produce localized and stable light pulses.  Such nonlinear synchronization processes where first observed shortly after the invention of the laser~\cite{didomenico1964small,hargrove1964locking,yariv1965internal,smith1970mode}.  
Modern ML lasers~\cite{haus2,duling} are now a mature, turn-key technology commonly used in many branches of science and commercial applications~\cite{siegman,weiner2011ultrafast}.  Traditionally, the key nonlinear process responsible for synchronizing cavity modes is a cubic Kerr nonlinearity, or intensity-dependent index of refraction.  Its interaction with linear dispersion and the cavity gain-loss dynamics are the basis of the canonical {\em master mode-locking equation}~\cite{haus3,haus4} which characterizes the equilibration of the pulse energy, and the consequent balance of nonlinearity and dispersion in forming stable pulses~\cite{kutz13}. 
More recently, spatio-temporal ML has been considered for creating light-bullet structures in Kerr media~\cite{kutz14, kutz15, kutz16,wright2017spatiotemporal}.  In this work, we develop a theory of spatio-temporal ML in quadratic media and show that stable ML can be achieved.  This provides an important extension of the ML theoretical framework to a broader class of problems whose quadratic nonlinearities can be leveraged with orders of magnitude less power.

Spatio-temporal ML is difficult to achieve in practice due to the physical balances that must be achieved in both the spatial and temporal domains.  And although ML has promoted a great deal of work in synchronizing temporal cavity modes, there has been relatively little work in understanding how to coherently superimpose spatial modes, or mode-lock in the spatial domain.  If both can be simultaneously synchronized, then spatio-temporal ML can be achieved.  The work of Wright et al.~\cite{wright2017spatiotemporal}, for instance, uses spatial filtering to achieve a variety of spatio-temporal ML states.  In many applications, however, the leading nonlinear polarization effects in an optical material is quadratic. 
Quadratic materials are referred to as $\chi^{2}$ materials, where$\chi^{2}$ is the second order susceptibility that describes second harmonic generation (SHG) first experimentally observed by Franken et al.~\cite{franken1961generation}.  Indeed, until such observations~\cite{armstrong1962interactions}, Maxwell's equations were thought to be linear.  Pulse shaping in quadratic media was proposed early on~\cite{ostrovskii}, with optical solitons theoretically predicted by Karamzin and Sukhorukov shortly after~\cite{karamzin1, karamzin2}.  In Belashenkov et al.~\cite{belashenkov} and DeSalvo et al.~\cite{desalvo}, experiments in a $\chi^{2}$ crystal demonstrated modulational instability and the self-defocusing phenomena typically observed in centro-symmetric $\chi^{3}$ materials. In 1995, optical solitons in a quadratic bulk material were observed by Torruellas et al.~\cite{torruellas1} , and existence of the solitons in $\chi^{2}$ waveguide were observed experimentally by Schiek et al.~\cite{schiek1}  in 1996. These original results have since been corroborated and extended in many follow-up experiments ~\cite{fuerst1,fuerst2,constantini1,courderc1,lopez1,bache1}, demonstrating that quadratic solitons can exist in both the spatial and the temporal domains in waveguides or bulk materials~\cite{torner1,buryak1, mihalache1,lutsky1}. One of the distinguished properties of quadratic nonlinear media is that it provides stable multidimensional pulse propagation without collapse in any dimension~\cite{torner1,buryak1, hayata1}.

It is well-known that the pulse dynamics in multidimensional non-resonant $\chi^{2}$  materials cannot be generally described by nonlinear Schr\"odinger (NLS) based equations~\cite{ablowitz1,ablowitz2, ablowitz3, ablowitz4}.  Indeed, these dynamics are governed by generalized NLS systems with coupling to a mean term (hereafter denoted as NLSM systems which are sometimes referred to as Benney--Roskes  or Davey--Stewartson type)~\cite{benney, davey}. 

Benney and Roskes~\cite{benney} first obtained NLSM equations in water of finite depth $h$ and without surface tension in 1969. In 1974, Davey and Stewartson~\cite{davey} reached an equivalent form of the NLSM equations by investigation of the evolution of a 3D wave packet in water of finite depth. The integrability of NLSM systems was studied in 1975 by Ablowitz and Haberman~\cite{haberman} in the shallow water limit. In 1977, the results of Benney and Roskes was extended to include surface tension by Djordevic and Reddekopp~\cite{djordevic}. Ablowitz et al.~\cite{ablowitz1, ablowitz2, ablowitz3} derived from first principles NLSM type equations describing the evolution of the electromagnetic field in quadratic nonlinear media. Recently, in
~\cite{bagci}, it was demonstrated that optical wave collapse can be arrested in the NLSM system by adding an external potential (lattice) to the model.

NLSM system is physically derived from an expansion of the slowly-varying wave amplitude in the first and second harmonics of the fundamental frequency and, a mean term that corresponds to the zeroth harmonic. This system describes the nonlocal-nonlinear coupling between a dynamic field that is related with the first harmonic and a static field that is related with the zeroth harmonic~\cite{ablowitz5}. The general NLSM system is given by~\cite{ablowitz2, ablowitz3,crasovan}
\begin{subeqnarray}
&& i{u_t} + {\nabla ^2}u + {\left| u \right|^2}u - \rho u\phi  = 0, \\
&& {\phi _{xx}} + \nu {\phi _{yy}} = {\left( {{\left| u \right|}^2} \right)_{xx}}
\label{eq:NLSM}
\end{subeqnarray}
where $u(x,t)$ is the normalized amplitude of the envelope of the normalized static electric field (which associated with the first-harmonic).  The parameter $\rho$ is a coupling constant that comes from the combined optical rectification and electro-optic effects  modeled by the $\phi(x,y)$ field, and $\nu$ is the coefficient that comes from the anisotropy of the material.

Given the long-history of the NLSM model and its broad applications to fluids and optics alike, we build on the pioneering optical work of Ablowitz and co-workers to characterize ML in quadratic nonlinear media.  Specifically, we modify the model to include the critical gain-loss dynamics that are a hallmark feature of ML systems.  Indeed, ML is manifestation of a broader class of damp-driven systems which are common across the sciences.  For instance, ML is also observed in rotating detonation engines (RDEs)~\cite{koch2020mode,koch2020multi}, where energy balances are similar to ML laser cavities~\cite{namiki1997energy,li2010geometrical}.  Thus we develop a ML theory for the NLSM optical systems, denoted ML-NLSM, by including bandwidth limited, saturating gain and cavity losses to model the overall ML dynamics which is capable of producing stable, 2D-soliton like solutions in the NLSM model.  A full stability analysis and bifurcation study is performed for this new ML-NLSM model.
Our ML-NLSM model is the quadratic, 2D analog of the master mode-locking theory of Haus.

The paper is outlined as follows: In Sec. 2, the ML-NLSM is presented and steady state solutions (fundamental solitons) of the model are obtained numerically. In Sec. 3, The mode-locking dynamics of the fundamental solitons are explored by direct numerical simulations of our derived ML-NLSM governing equations.  Section~4 provides a stability analysis which details the
linear stability of the fundamental spatio-temporal solitons. Results of the study is discussed in Sec. 5.
\section{(2+1)D NLSM Systems}
Our theoretical considerations begin by considering the NLSM model~\cite{ablowitz2, ablowitz3,crasovan} in Eq.~(\ref{eq:NLSM}).  The model is modified to account for cavity losses and a bandwidth limited, saturating gain term which is canonical in ML models for lasers~\cite{kutz13}.
The ML-NLSM model in (2+1) dimensions is given by
\begin{subeqnarray} \label{NLSM2}
&& i{{u}_t} + \frac{D}{2} {\nabla^2 u}+  {\beta\left| {u} \right|^2}{u} -\rho \phi u=iRu, \\
&& \phi_{xx}+\nu \phi_{yy}=(|{u}|^2)_{xx} 
\end{subeqnarray}
where $R$ is the gain-loss operator given by
\begin{equation} 
R=g(t)(1+\tau\nabla^2)-\gamma_0 -{p\left| {u} \right|^4} +\alpha\phi
\end{equation}
with the time-depended gain saturation dynamics $g(t)$ given by
\begin{equation}
g(t)= \frac{2g_0}{1+||u||^2/e_0}.
\end{equation}
In the formulation, $u(x,y,t)$ is a function of time $t$ and the transverse variables $x$ and $y$. $D$ denotes the average diffraction coefficient. The evolution dynamics is coupled to the $\phi(x,y)$ field. The parameter $\beta$ represents the strength of the cubic nonlinearity and $p$ represents the strength of the quintic self-phase modulation term. The coupling parameter $\rho$  describes the combined optical rectification and electro-optic effects. In the operator $R$, all parameters are positive. These include the gain bandwidth $\tau$ and the linear attenuation parameter $\gamma_0$. The dynamic gain $g(t)$ depends on the input pump strength $g_0>0$, cavity saturation energy $e_0$,  and the total cavity energy ($L^2$-norm) ${\left\| u \right\|^2} = {\int\!\!\int {{{\left| u \right|}^2}dxdy}}$ where integration is performed over the entire space of $x$ and $y$~\cite{kutz16}.

The ML-NLSM equation (\ref{NLSM2}) along with its solutions and dynamics represents the primary contribution of this manuscript.  The model has a number of features of note.  First, the model is a (2+1) dimensional model in electric field envelop $u(x,t)$.  For $\rho=0$ and $R=0$ the (2+1) NLS equation has solutions that exhibit finite time blow-up of soluions~\cite{landman1988rate}.  This behavior is regularized by both the coupling to $\phi$ and the gain-loss dynamics given by the operator $R$.  Importantly, this model includes the material anisotropy through the parameter $\nu$.  Of primary interest is to determine if localized ML solutions exist for this system, thus allowing the ML-NLSM to support quadratic spatio-temporal solitons.  Such soliton-like solutions are considered  in the following subsections using recently developed numerical methods for finding steady-state solutions of partial differential equations.

\subsection{Numerical Solutions using the Squared Operator Method (SOM)}
In order to obtain the fundamental soliton solutions of the (2+1)D ML-NLSM, we use a modification of the computational algorithm called the {\em squared operator method} (SOM)~\cite{yang}. The method is based on iterating new differential equations whose linearization operators are squares of those of the original equations, together with an acceleration technique. The scheme of the method is outlined in what follows.

Substituting the ansatz $u = U\left( {x,y} \right){\exp ({i\mu t}})$ into the model (\ref{NLSM2}), we get the operator
\begin{eqnarray}
\begin{array}{c}
{{\bf{L}}_0}{\bf{u}} = \frac{D}{2}{\nabla ^2}U + \beta {\left| U \right|^2}U + i\gamma U - i{g(t)}(1 + \tau {\nabla ^2}) \\
+ ip{\left| U \right|^4}U - \rho \phi U - i\alpha \phi U - \mu U,\\ \\
{\phi _{xx}} + \nu {\phi _{yy}} = {\left( {{{\left| U \right|}^2}} \right)_{xx}}
\end{array}
\end{eqnarray}
where $\mu$ is propagation constant. Separating the operator ${{\bf{L}}_0}$ into its real and imaginary parts gives the following sub-operators,
\begin{eqnarray}
\begin{array}{c}
T1 = {{\rm Re}} \left( {{{\cal F}^{ - 1}}\left( {\frac{{{\cal F}\left( {{{\bf{L}}_0}{\bf{u}}} \right)}}{{{K^2} + c}}} \right)} \right),\\ \\
T2 = {{\rm Im}} \left( {{{\cal F}^{ - 1}}\left( {\frac{{{\cal F}\left( {{{\bf{L}}_0}{\bf{u}}} \right)}}{{{K^2} + c}}} \right)} \right).
\end{array}
\end{eqnarray}
\noindent where $\cal F$ symbolize Fourier transformation, $k=(k_x, k_y)$ are wavenumbers in the $x$ and $y$ directions respectively, $K^2=k_x ^2+k_y ^2$ and $c$ is parameter for parametrizing the numerical scheme.  Decomposing the amplitude into its real and imaginary parts $U = u(x,y) + iv(x,y)$ and inserting into ${{\bf{L}}_0}{\bf{u}}$, we get the two sub-operators
\begin{eqnarray}
\begin{array}{c}
{L_{{{\rm Re}}}} = \frac{D}{2}{\nabla ^2}u+\beta ({u^3} + u{v^2})-\gamma v+{g(t)}(1+\tau{\nabla ^2})v \\  -p\left( {{u^4}v + 2{u^2}{v^3} + {v^5}} \right) - \rho \phi u + \alpha \phi v - \mu u \\
\\
{L_{{{\rm Im}}}} = \frac{D}{2}{\nabla ^2}v + \beta ({v^3} + {u^2}v) - \gamma u - {g(t)}(1 + \tau {\nabla ^2})u  \\  + p\left( {{u^5} + 2{u^3}{v^2} + u{v^4}} \right) - \rho \phi v - \alpha \phi u - \mu v .
\end{array}
\end{eqnarray}
Taking partial derivatives of these new operators with respect to both $u$ and $v$ gives the matrix components
\begin{eqnarray}
\begin{array}{c}
{R_{11}} = \frac{{\partial {L_{{\mathop{\rm Re}\nolimits} }}}}{{\partial u}}(T1)\,,\, \, {R_{12}} = \frac{{\partial {L_{{\mathop{\rm Re}\nolimits} }}}}{{\partial v}}(T2),\\ \\  {R_{21}} = \frac{{\partial {L_{{\mathop{\rm Im}\nolimits} }}}}{{\partial u}}(T1)\,,\,\,  {R_{22}} = \frac{{\partial {L_{{\mathop{\rm Im}\nolimits} }}}}{{\partial v}}(T2) .
\end{array}
\end{eqnarray}
Using the elements of this matrix, the final operator is defined as
\begin{equation}
{{\bf{L}}_1}{\bf{u}} = {R_{11}} + {R_{12}} + i({R_{21}} + {R_{22}}).
\end{equation}
After the operator is defined, the iteration scheme is implemented as follows,
\begin{eqnarray}
\begin{array}{l}
{U_{n + 1}} = {U_n} - \left( {{{\cal F}^{ - 1}}\left( {\frac{{{\cal F}\left( {{{\bf{L}}_1}{\bf{u}}} \right)}}{{{K^2} + c}}} \right)} \right)\Delta t\,,\\ \\
{\mu _{n + 1}} = {\mu _n} + \|{u \cdot T1 + v \cdot T2} \|\Delta t, \\ \\
{\phi _{n + 1}} = {{\cal F}^{ - 1}}\left( {\frac{{k_x^2{\cal F}\left( {{{\left| {{U_n}} \right|}^2}} \right)}}{{k_x^2 + \nu k_y^2}}} \right) .
\end{array}
\end{eqnarray}
This numerical algorithm is iterated from an initial guess until the error $ E=\sqrt {\| U_{n + 1} - {U_n} \|^2}  + \left| {{\mu _{n + 1}} - {\mu _n}} \right| < 10^{-6} $.  This algorithm has been demonstrated to be efficient and accurate in computing localized solutions for a wide-range of nonlinear PDEs~\cite{yang}.  It is also effective for the ML-NLSM model proposed here in generating the desired mode-locked spatio-temporal states of interest.
\subsection{Numerical Existence of the Fundamental Solitons}
\begin{figure}[t]
  \hspace*{-.2in}
  \includegraphics[scale=0.52]{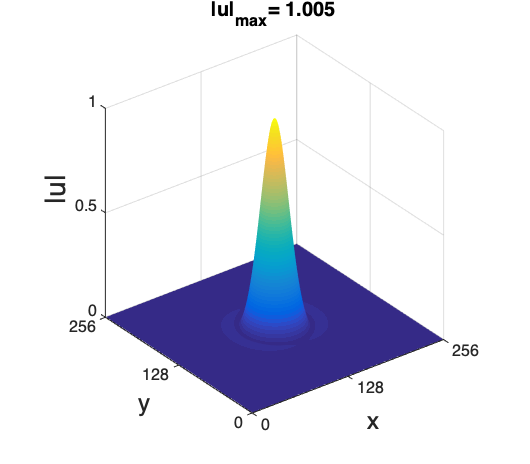}
   \caption{Fundamental soliton that is obtained when $\rho=0.5$, $\nu=1$, $c=5$ and $\Delta t=0.1$.}
  \label{fundamental1}
\end{figure}

The SOM algorithm is used to compute a steady state solution (fundamental soliton) of the ML-NLSM Eq.~(\ref{NLSM2}). Once the solution is obtained, it can be used to the linear stability properties of the solitons.
In what follows, the following set of base parameters are used.  Unless otherwise specified, we set
\begin{equation}\label {parameter1}
\left( {D,\beta ,\gamma ,{g_0},{E_0},\tau ,p,\alpha ,\mu } \right) = (1,1,1,4.88,1,0.08,0.5,1,1). 
\end{equation}
With these parameters, the numerical convergence to the fundamental soliton (steady state solution) is shown in Fig.~\ref{fundamental1} for the parameter values $\rho=0.5$ and $\nu=1$ in the ML-NLSM system when $c=5$ and $\Delta t=0.1$ in the algorithm. In addition, we have found that the fundamental soliton solution can be obtained for $0 \le \rho \le 1.7$ when $\nu=1$ with suitable $c$ and $\Delta t$ parameters.
\begin{figure*}[t]
  \hspace*{-.4in}
  \includegraphics[width=1.1\textwidth]{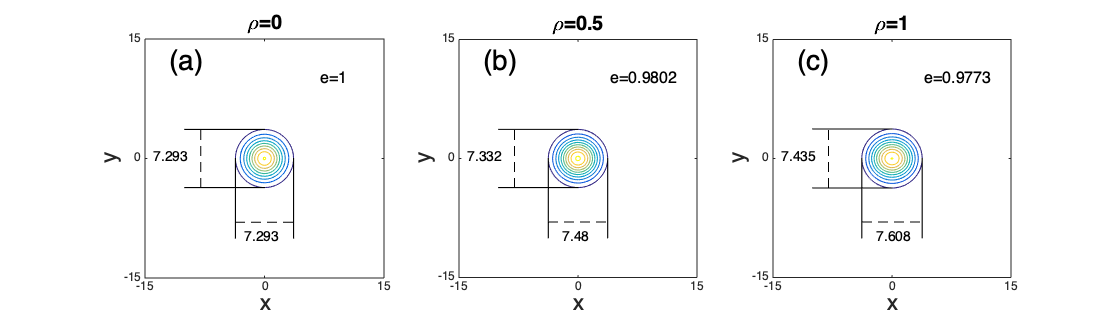}
\caption{Contour image of the fundamental soliton for (a)$\rho = 0$; (b) $\rho = 0.5$; (c) $\rho = 1$. All solitons are obtained when $\nu=1$.}
\label{astigmat1}
\end{figure*}

It is noteworthy that, as demonstrated in previous studies~\cite{bagci, ablowitz5}, due to the anisotropy of the  ML-NLSM system, steady state solutions do not possess radial symmetry. In the other words, the ML-NLSM model generates astigmatic fundamental solitons. To explore the level of astigmatism in the solitions, we define
\begin{equation}
e=\frac{\mbox{radius along $y$-axis}} {\mbox{radius along $x$-axis}}
\end{equation}
as a measure of astigmatism. When $e = 1$, the solution corresponds to a radially-symmetric fundamental soliton, and $e < 1$ and $e >1$ correspond to a soliton that is relatively wider along the $x$ and $y$ axes, respectively.  Thus it takes on an elliptical shape for $e\neq 1$.

Contour images of fundamental solitons are plotted in Fig.~\ref{astigmat1} for $\rho=0$, $\rho=0.5$ and $\rho=1$, respectively. It can be seen from Fig.~\ref{astigmat1} that as $\rho$ increases, the contours of the fundamental soliton become more astigmatic along the $x$-axis. On the other hand, we observe that as the anisotropy coefficient $\nu$ increases (for a fixed $\rho$), the fundamental soliton become less astigmatic along $x$-axis and after a threshold value of $\nu$ (i.e., when $\nu \ge 1.3$ and $\rho=0.5$) the solitons become relatively wider along $y$-axis.

\section{Dynamics of Fundamental Soliton}

\begin{figure*}[t]
    \hspace*{-.4in}
  \includegraphics[width=1.0\textwidth]{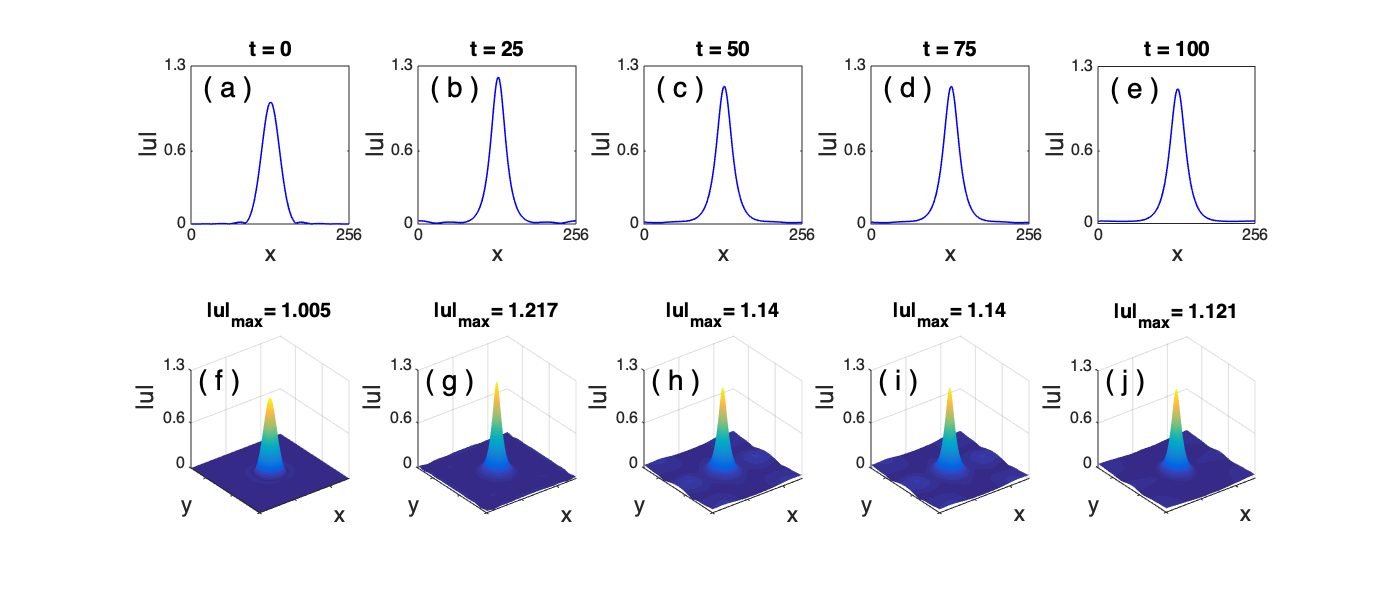}
   \caption{Nonlinear evolution of the fundamental soliton from $t=0$ to $t=100$. Five snapshots of the soliton captured at different propagation times. Profile of the evolved soliton along x-axis is plotted (upper panels) while (a) $t=0$, (b) $t=25$, (c) $t=50$, (d) $t=75$, (e) $t=100$, and corresponding 3D view of the evolved soliton is plotted (lower panels) while (f) $t=0$, (g) $t=25$, (h) $t=50$, (i) $t=75$ and (j) $t=100$. The fundamental soliton is obtained when $\rho=0.5$, $\nu=1$ and $\alpha=1$.}
  \label{nonlinear1}
\end{figure*}

To investigate the dynamics of the ML-NLSM solitons, we directly simulate Eq.~(\ref{NLSM2}) 
for long times.  A finite-difference discretization scheme is used in the spatial domain and the solution is advanced in time with a fourth-order Runge-Kutta method.
We plot 3D views and profiles of the solitons versus the propagation time during the evolution (from $t=0$ to $t=t_{max}$). Stable (mode-locked) soliton solutions should nearly preserve its mode shape, profile and peak amplitude over time.
Fig.~\ref{nonlinear1} shows the evolution of the fundamental soliton (obtained in Fig.~\ref{fundamental1}) from $t\in[0,100]$ with a numerical time-step of $dt=0.001$.  Snapshots of the evolution dynamics are plotted for $t=0$, $t=25$, $t=50$, $t=75$ and $t=100$.

\begin{figure*}[t]
      \hspace*{-.4in}
  \includegraphics[width=1.0\textwidth]{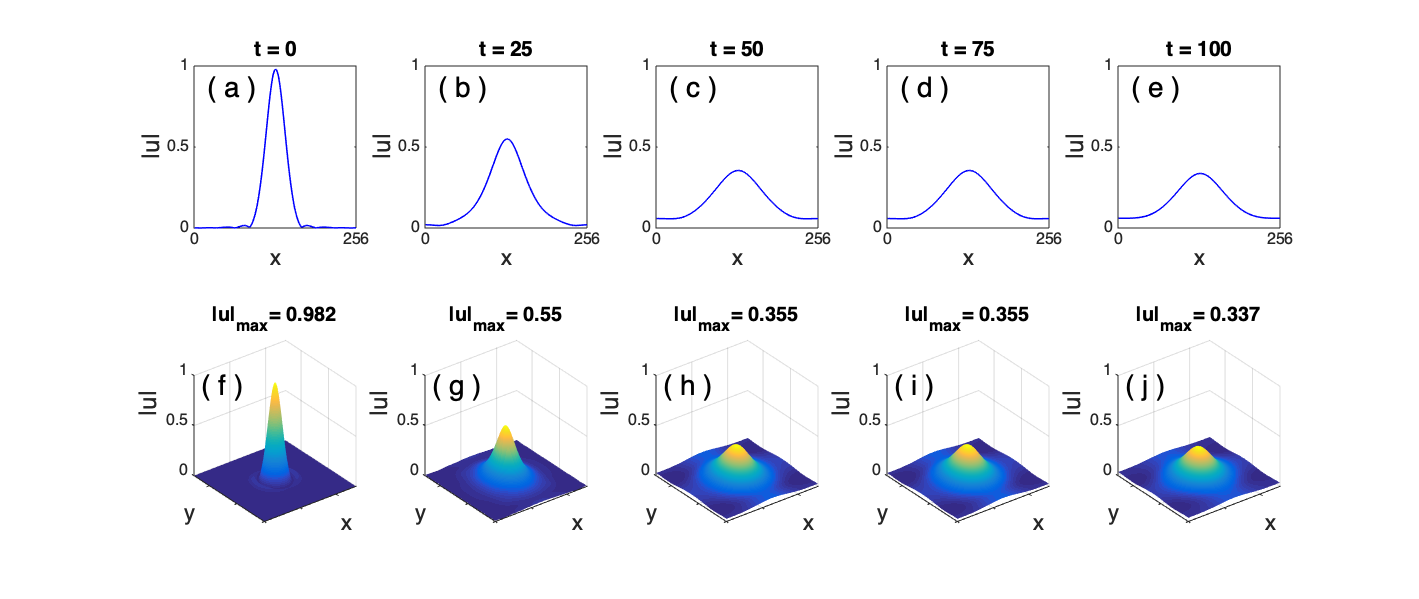}
     \caption{Collapse of the fundamental soliton, that is obtained when $\rho=1$, $\nu=1$ and $\alpha=1$, is demonstrated with five snapshots of the soliton captured at different propagation times. Profile of the evolved soliton along x-axis is plotted (upper panels) while (a) $t=0$, (b) $t=25$, (c) $t=50$, (d) $t=75$, (e) $t=100$, and corresponding 3D view of the evolved soliton is plotted (lower panels) while (f) $t=0$, (g) $t=25$, (h) $t=50$, (i) $t=75$ and (j) $t=100$.}
  \label{nonlinear2}
\end{figure*}

As can be seen from the Fig.~\ref{nonlinear1}, the profile of the evolved soliton (upper panels) is preserved and the peak amplitude of the fundamental soliton oscillates with relatively small amplitude during the evolution.  At the end of the simulation for $t=100$, the mode shape of the evolved soliton (lower panels) is shown to be preserved. These results demonstrate that the considered soliton is mode-locked in this parameter regime.

The fundamental solitons are mode-locked when $0\le \rho < 0.8$, $\nu=1$ and $\alpha=1$ in the ML-NLSM model. When $\rho \ge 0.8$,  the peak amplitude of the fundamental solitons decay after a short time of evolution and the solitons collapses and is not self-supporting.  The collapse of the fundamental soliton computed for $\rho=1$, $\nu=1$ and $\alpha=1$, is plotted in Fig.~\ref{nonlinear2}. It is obvious that the soliton can not stay mode-locked during the evolution since the amplitude of the soliton decreases significantly during the evolution (see Fig.~\ref{nonlinear2} upper panels) and the mode shape is no longer preserved during evolution (see Fig.~\ref{nonlinear2} lower panels).

\begin{figure*}[t]
      \hspace*{-.4in}
  \includegraphics[width=1.0\textwidth]{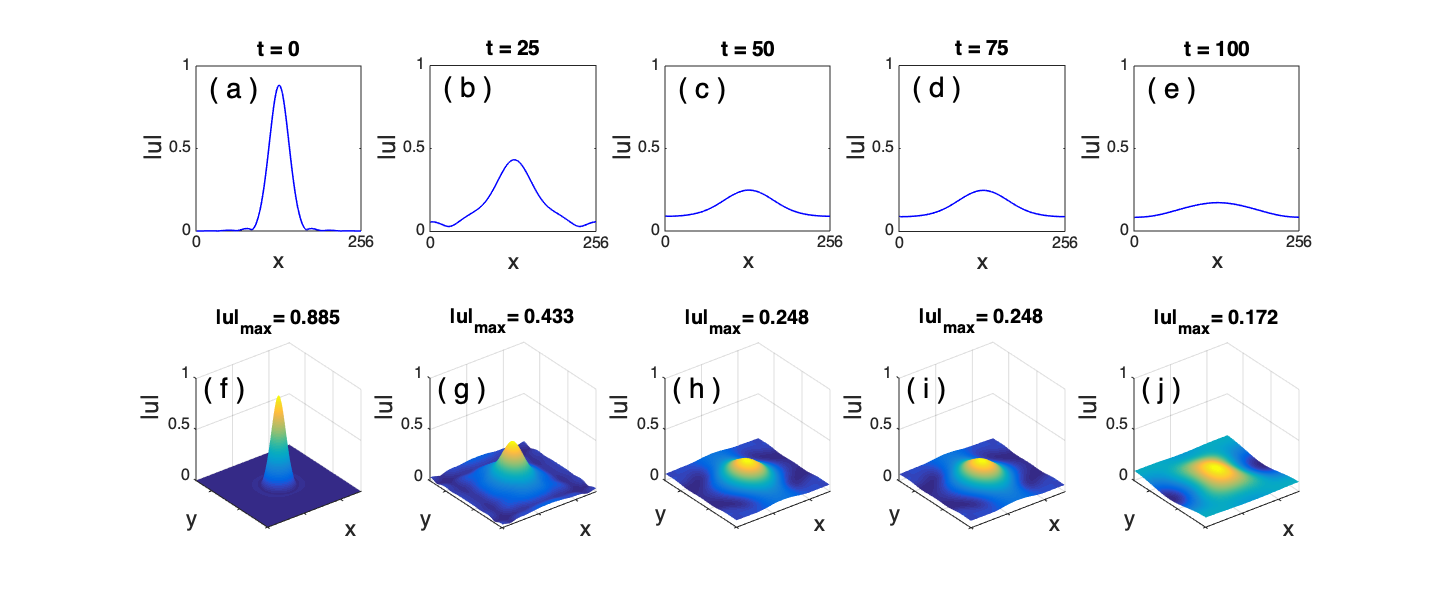}
       \caption{Collapse of the fundamental soliton, that is obtained when $\rho=0.5$, $\nu=1$ and $\alpha=0$, is demonstrated with five snapshots of the soliton captured at different propagation times. Profile of the evolved soliton along x-axis is plotted (upper panels) while (a) $t=0$, (b) $t=25$, (c) $t=50$, (d) $t=75$, (e) $t=100$, and corresponding 3D view of the evolved soliton is plotted (lower panels) while (f) $t=0$, (g) $t=25$, (h) $t=50$, (i) $t=75$ and (j) $t=100$.}
  \label{nonlinear3}
\end{figure*}

In addition, it has seen that $\alpha$ parameter (which shows quadratic polarization effect) plays an important role in mode-lock of the fundamental solitons. Fundamental solitons that are obtained when $\rho=0.5$ and $\nu=1$ are stable for $0.6<\alpha<1.8$.
While for $\alpha=0$ and $\rho>0$ in the ML-NLSM model, the fundamental solitons collapse and no mode-locking occurs.  Collapse of the fundamental soliton for $\rho=0.5$, $\nu=1$ and $\alpha=0$ is showed in Fig.~\ref{nonlinear3}. One can easily see that peak amplitude of the soliton decreases sharply after a short propagation distance (see Fig.~\ref{nonlinear3} upper panels) and the soliton is dispersed away entirely during the evolution (see Fig.~\ref{nonlinear3} lower panels).
 
\begin{figure*}[t]
  \begin{center}
  \center{\scalebox{0.8}{\includegraphics*{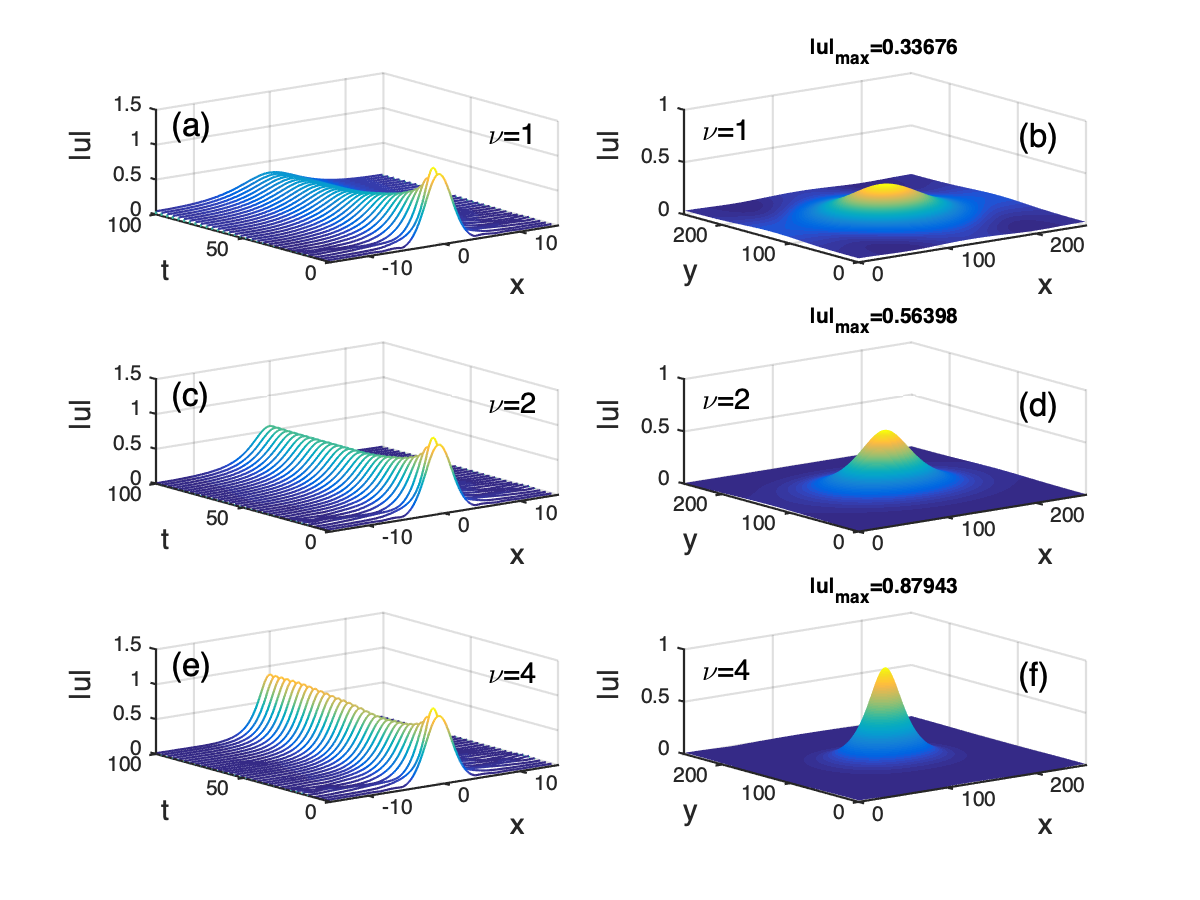}}}
   \caption{Nonlinear evolution of the fundamental soliton for larger anisotropy coefficients  (a) when $\nu=1$;  (c) when $\nu=2$  and (e) when $\nu=4$ . 3D view of the soliton after evolution at $t=100$ (b) when $\nu=1$;  (d) when $\nu=2$  and (f) when  $\nu=4$. All fundamental solitons are obtained for $\rho=1$ and $\alpha=1.$}
  \label{compare_nu}
  \end{center}
\end{figure*}

Furthermore, we have seen that the collapse of the fundamental solitons can be arrested (or delayed) by increasing the value of the anisotropy coefficient $\nu$. As a special case we have increased the value of $\nu$ from 1 to 4 and plot the evolution of the fundamental soliton that is obtained for $\rho=1$ and $\alpha=1$ in Fig.~\ref{compare_nu}.  It is clearly seen that although the soliton, which is obtained for $\nu=1$, does not blow up in finite time, it cannot be considered as a robust since the amplitude of the soliton decreases significantly  after $t=10$ (see Fig.~\ref{compare_nu}(a)) and finally (at $t=100$) it decays (see Fig.~\ref{compare_nu}(b)). To improve the stability of considered soliton, we increase the value of $\nu$ to $\nu=2$ and $\nu=4$ and depict the nonlinear evolution of the soliton for each case in Fig.~\ref{compare_nu}(c) and (e), respectively. We observe that the collapse of the soliton is delayed when $\nu=2$ (see Fig.~\ref{compare_nu}(c) and (d)) and, collapse of the soliton is prevented when $\nu=4$ (see Fig.~\ref{compare_nu}(e) and (f)).

It should be pointed out that, in real optical systems, increasing the anisotropy parameter $\nu$, may not be used as a collapse arrest mechanism in some cases, since $\rho$ and $\nu$ parameters are fixed values depending on the type of material that is considered. However, the anisotropy $\nu$ can be applied to improve the stability of solitons in the range of real physical parameter regime. 
\section{Linear Stability Analysis}
A standard way for determining stability is to calculate the spectrum of linearization of the model (\ref{NLSM2}) about the fundamental soliton solutions computed with the SOM technique. By denoting 
\begin{equation}
u=e^{-i\theta t}[u_0(x,y)+\tilde u(x,y,t)] 
\end{equation}
where $u_0(x,y)$ is the fundamental soliton, $\theta$ is propagation constant and $\tilde{u}\ll 1$ is the infinitesimal perturbation. If the perturbation $\tilde{u}$ decays to zero, then the fundamental soliton is considered to be linearly stable. Inserting the perturbed solution into the equation (\ref{NLSM2}), we get the linearized system for $\tilde u$ by neglecting small terms of the second order $O(\tilde u ^2)$: 
\begin{eqnarray}\label{linear1}
 \begin{array}{c}
\tilde{u}_t=i\theta \tilde{u}+i\frac{D}{2}{\nabla ^2}\tilde{u} + i\beta (2{\left| u_0 \right|^2}\tilde{u} \!+\!u_0^2\tilde{u}^*)-\gamma \tilde{u} \\ \\ \!+\! {g(t)}(1 \!+\! \tau {\nabla ^2})\tilde{u} - p(3{\left| u_0 \right|^4}\tilde{u}+2{\left| u_0 \right|^2}u_0^2\tilde{u}^*)  \\ \\ - i\rho \phi\tilde{u}+\alpha \phi\tilde{u}.
 \end{array}
\end{eqnarray}
Separating the fundamental soliton and the perturbations into real and imaginary parts as follows
\begin{equation}
u_0=a_0+ib_0,\,\,\,\,\, {\tilde{u} = R_0e^{\lambda t}+iI_0e^{\lambda t}},\,\,\,\,\, 
\end{equation}
we obtain $\tilde{u}_t=\lambda \tilde{u}$,
and substituting $u_0$, $\tilde{u}$ into the system (\ref{linear1}) results in the eigenvalue problem
\begin{equation}
  {\bf AV}=\lambda\bf V
\end{equation}
where 
\begin{equation*}
A= 
\begin{pmatrix}
F_R & G_I  \\
G_R & F_I 
\end{pmatrix},
\,\,\,\,\,\,\,\,V= 
\begin{pmatrix}
R_0 \\
I_0
\end{pmatrix}.
\end{equation*}
If the real part of the $\lambda$ is positive,  the fundamental soliton is unstable. The eigenvalues of $\bf A$ can be calculated numerically with finite difference discretization of the spatial domain.  Note that the matrix coefficients of ${\bf A}$ are given by
\begin{eqnarray}
 \begin{array}{c}
F_R=-2\beta a_0 b_0-\gamma_0+g(t)(1 + \tau {\nabla ^2})-p(5a_0^4+b_0^4+6a_0^2b_0^2) \\ +\alpha\phi_0 - \frac{4g_0}{(1+||u||^2)^2}(1 + \tau {\nabla ^2}){a_0} \** a_0, \\  \\
G_I= -\left(\frac{D}{2}{\nabla ^2} + \beta (a_0^2 +3b_0^2)+4p(a_0^3b_0+a_0b_0^3)+\theta-\rho\phi\right) \\ -\frac{4g_0}{(1+||u||^2)^2}(1 + \tau {\nabla ^2}){a_0}\** b_0, \\  \\
F_I=2\beta a_0 b_0-\gamma_0+g(t)(1 + \tau {\nabla ^2})-p(a_0^4+5b_0^4+6a_0^2b_0^2) \\ +\alpha\phi_0 
-\frac{4g_0}{(1+||u||^2)^2}(1 + \tau {\nabla ^2}){b_0}\** b_0,\\  \\
G_R= \left(\frac{D}{2}{\nabla ^2} + \beta (3a_0^2 +b_0^2)-4p(a_0^3b_0+a_0b_0^3)+\theta-\rho\phi\right)\\-\frac{4g_0}{(1+||u||^2)^2}(1 + \tau {\nabla ^2}){a_0} \** b_0.
 \end{array}
\end{eqnarray}
The $\**$ notation denotes the integral $a_0\** b_0\!=\!\int_{-\infty}^{\infty}{a_0(\tau) b_0(\tau) d\tau}$, which results from the nonlocal behavior given by the saturated gain dynamics~\cite{farnum}.

\begin{figure*}[t]
      \hspace*{-.4in}
  \includegraphics[width=1.1\textwidth]{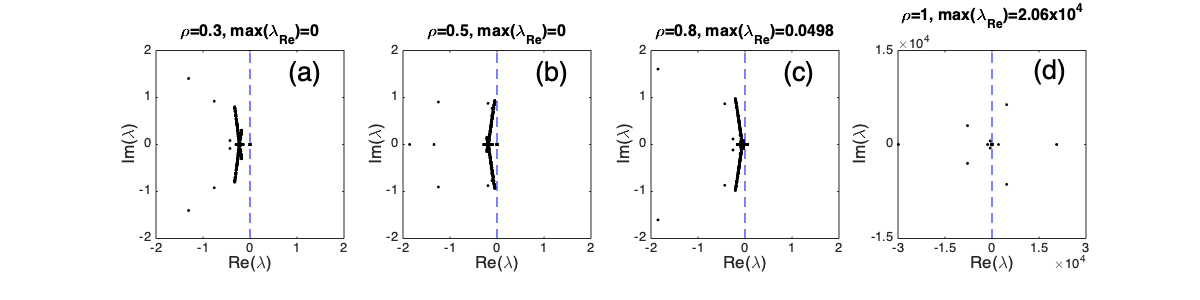}
   \caption{\label{spectrum1} Real and imaginary part of the numerically computed eigenvalues of operator $\bf A$ when the fundamental solitons ($u_0$) are obtained for (a)$\rho=0.3$, (b)$\rho=0.5$, (c)$\rho=0.8$, (d)$\rho=1$.}
\end{figure*}

The linear spectra of fundamental solitons can be computed by evaluating the matrix ${\bf A}$.  The spectra obtained for $\rho=0.3$, $\rho=0.5$, $\rho=0.8$ and $\rho=1$ are plotted in Fig.~\ref{spectrum1}. It can be seen that the fundamental soliton for $\rho=0.3$ and $\rho=0.5$ has an eigenvalue spectra whose have a positive real parts $Re(\lambda)$ (see Fig.~\ref{spectrum1}(a)-(b)) are in the left-half plane, thus showing these ML-NLSM solitons to be linear stable. For $\rho=0.8$ and $\rho=1$ (see Fig.~\ref{spectrum1}(c)-(d)) a number of  eigenvalues have positive real part which indicates that these fundamental solitons are linearly unstable. 

In addition, we have seen that the fundamental solitons' spectra include eigenvalues with positive real part when $\rho \ge 0.8$ which demonstrate the existence of a unstable soltion region for the given parameter regime (\ref{parameter1}). Similar to being nonlinearly unstable, the fundamental soltions become linearly unstable when $\alpha=0$ and $\rho>0$.
\section{Conclusion}
The proposed ML-NLSM model has been formulated as an extension of the master mode-locking model by the addition of higher-order dispersion and quadratic electro-optic effects.  This study reveals the potential of using quadratic nonlinear media to generate spatio-temporal mode-locked soliton states in a nonlinear optical system.  Using modern computational methods we have shown that there exists steady state soliton solutions of the ML-NLSM mode-locking model that are astigmatic in nature. Stability of the ML-NLSM states have been characterized by direct numerical simulation of the model as well as by linear stability arguments and computation of the spectra of the linearized operator.  Both show that the soliton solutions of the ML-NSLM model have regions of stable spatio-temporal mode-locking.

Specifically, It has been shown that when the coupling constant $\rho$ (that comes from the combined optical rectification and electro-optic effects) is smaller than 0.8, there is no eigenvalue with a positive real part in the spectrum of linearization of the ML-NLSM model, thus showing that the fundamental solitons in this region are stable, mode-locking states that act as attractors.
The nonlinear stability of the fundamental solitons have been examined by direct simulations of the ML-NLSM model and the results demonstrate that for $\rho<0.8$, the fundamental solitons' profile are preserved and the peak amplitude of the solitons oscillates relatively small amplitude during the nonlinear evolution which means stable mode-locking is achieved during the evolution.

In addition, it has seen that $\alpha$ parameter (which shows quadratic polarization effect) has an critical importance for mode-locking operation in the ML-NLSM model. Specifically for $\alpha=0$  and $\rho>0$, the fundamental solitons collapse and none of the ML-NLSM soliton states stay mode-locked.  
In conclusion, we have constructed the ML-NLSM model as a modification of master mode-locking model and demonstrated the possibility of mode-locking of astigmatic steady-state solutions in the quadratic nonlinear media.
\begin{acknowledgments}
The first author was supported by TUBITAK 2219-International Postdoctoral Research Fellowship Programme under Grant No.~1059B191600743.
\end{acknowledgments}

\bibliography{nlsm}

\end{document}